\documentclass[conference]{IEEEtran}
\usepackage{fancybox}
\usepackage[ruled,vlined]{algorithm2e}
\usepackage{graphicx}
\usepackage{paralist}
\usepackage{tabularx}
\usepackage[hyphens]{url}
\usepackage{balance}
\usepackage{multirow}
\usepackage{multicol}
\usepackage{pbox}
\usepackage{mathtools}
\usepackage{cite}

\usepackage[TABBOTCAP]{subfigure}
\usepackage{url}
\usepackage{balance}
\usepackage{amsmath}
\usepackage{verbatim}
\usepackage{float}
\usepackage{colortbl}
\usepackage{amssymb}
\usepackage{ifthen}
\usepackage{pifont}
\include{macro}

\newcommand{\monkeylabsp}{{\sc MonkeyLab}}
\newcommand{\monkeylab}{{\sc MonkeyLab~}}

\newcommand\blfootnote[1]{%
  \begingroup
  \renewcommand\thefootnote{}\footnote{#1}%
  \addtocounter{footnote}{-1}%
  \endgroup
}

\begin{document}
\clubpenalty=10000
\widowpenalty = 10000000
\displaywidowpenalty = 1000000

\pagenumbering{arabic}
\pagestyle{plain}
\newenvironment{myquote}{\list{}{\leftmargin=0.05in\rightmargin=0.05in}\item[]}{\endlist}

\setlength{\floatsep}{2pt}

\setlength{\textfloatsep}{1pt}

\title{Mining Android App Usages for Generating Actionable GUI-based Execution Scenarios}

\author{
\IEEEauthorblockN{Mario Linares-V\'asquez, Martin White, Carlos Bernal-C\'ardenas, Kevin Moran, and Denys Poshyvanyk}
\IEEEauthorblockA{The College of William and Mary, Williamsburg, VA, USA \\
\{mlinarev, mgwhite, cebernal, kpmoran, denys\}@cs.wm.edu 
} 
\vspace{-1cm}
}
\maketitle
\thispagestyle{plain}
\begin{abstract}
GUI-based models extracted from Android app execution traces, events, or source code can be extremely useful for challenging tasks such as the generation of scenarios or test cases. However, extracting effective models can be an expensive process. Moreover, existing approaches for automatically deriving GUI-based models are not able to generate scenarios that include events which were not observed in execution (nor event) traces. In this paper, we address these and other major challenges in our novel hybrid approach, coined as \monkeylabsp. 
Our approach is based on the {\tt \textit{Record}}$\rightarrow${\tt \textit{Mine}}$\rightarrow${\tt \textit{Generate}}$\rightarrow${\tt \textit{Validate}} framework, which relies on recording app usages that yield execution (event) traces, mining those event traces and generating execution scenarios using statistical language modeling, static and dynamic analyses, and validating the resulting scenarios using an interactive execution of the app on a real device.
The framework aims at mining models capable of generating feasible and fully replayable (i.e., actionable) scenarios reflecting either natural user behavior or uncommon usages (e.g., corner cases) for a given app. We evaluated \monkeylab in a case study involving several medium-to-large open-source Android apps. Our results demonstrate that \monkeylab is able to mine GUI-based models that can be used to generate actionable execution scenarios for both natural and unnatural sequences of events on Google Nexus 7 tablets. 

\end{abstract}


\section{Introduction}
\label{sec:intro}
The mobile handset industry has been growing at an unprecedented 23\% compound annual growth rate in revenue since 2009 \cite{Vision:2013}, and the growth between 2012 and 2016 is expected to be around 28\% \cite{VisionMobile:Q32013}. This global app economy with millions of apps (1.3M+ Android and 1.2M+ iOS apps), 2.3M developers, and billions of devices and users has been a tremendous success \cite{VisionMobile:14}. Many of these mobile apps have features that rival their desktop counterparts and belong to several domain categories spanning from games to medical apps. Mobile platforms enable user interaction via touchscreens and sensors (e.g., accelerometer, ambient temperature, gyroscope) that present new challenges for software testing. Furthermore, mobile developers and testers face other emerging challenges such as rapid platform/library evolution and API instability \cite{Linares-Vasquez:FSE13, Linares:ICPC14, ICSM13:KIM,Bavota:TSE15}, platform fragmentation \cite{Han:WCRE2012}, continuous pressure from the market for frequent releases \cite{Jones:2014,Hu:EuroSys14}, and limited availability/adequacy of testing tools for mobile apps \cite{Mona:ESEM13, Nagappan:FSE14}, among many other challenges. 

Although several tools are available to support automated execution of Android apps for validation purposes, in practice, testing is still performed mostly manually as recent survey study results show  \cite{Mona:ESEM13}.   Limited testing time during the development process tends to prohibit the use of data from these manually generated scenarios for different devices \cite{Mona:ESEM13} in a continuous development fashion, especially taking into account that some apps may have users coming from as many as 132 unique devices \cite{Nagappan:FSE14}.  A significant  amount of work is required to generate replay scripts that are coupled to screen dimensions for a single device (i.e., one script is required for each target device). Consequently, these scripts become quickly outdated when significant changes are made to the app's GUI \cite{Mona:ESEM13, GRECHANIK:ICSM09}. 

Existing research tackled some of these issues by deriving models that represent the GUI and behavior of apps. These models are abstract representations that can be decoupled from device dimensions and event locations and can still remain valid when small changes are done to the app (e.g., button location change). For instance, some representative approaches for deriving such models use either dynamic \cite{Takala:ICST11, Ravindranath:MobiSys14, Amalfitano:ASE12, Azim:OOPSLA13, Machiry:FSE13, Choi:OOPSLA13, Nguyen:TSE14, Tonella:ICSE14} or static analyses~\cite{Rountev:CGO14,Mirzaei:SIGSOFT12,Jensen:ISSTA13,Yeh:SERE-C14, Anand:FSE12,Elbaum:TSE2005}. However, current approaches fall short in (i)  generating scenarios that are representative of natural (i.e., typical end-user) application usages \cite{Tonella:ICSE14}, (ii) taking into account the context which exists in an app's execution history \cite{Dias:WEASELTech7, Tonella:ICSE14}, (iii) generating  sequences with previously unseen (i.e., not available in artifacts used to derive the model) but feasible events. Moreover, utilizing model-based testing techniques for GUI-based testing in industrial contexts may be particularly challenging because creating such models requires specialized expertise and using these models (for manual and automated testing) assumes a logical mapping between the model and the actual system that was modeled \cite{Aho:ICSTW14}. 
\blfootnote{*This work is supported by NSF CCF-1218129 and CCF-1253837 grants.
grants. Any opinions, findings, and conclusions expressed herein are
the authors and do not necessarily reflect those of the sponsors.}

In practice, developers constantly test their apps \textit{manually} by exercising apps on target devices. In fact, manual testing is usually preferred over automated approaches for testing mobile apps \cite{Mona:ESEM13}. However, oftentimes this execution data is simply thrown on the ground and never used when an app needs to be retested. \textit{Our key hypothesis is that all this data that is produced from regular app usages by developers, testers, or even end-users can be effectively recorded and mined to generate representative app usage scenarios (as well as the corner cases) that can be useful for automated validation purposes.} Furthermore, our intuition is that the models mined from execution (event) traces can be augmented with static analysis information to include unseen but feasible events.

In this paper, we propose a novel hybrid approach for mining GUI-based models from event logs collected during routine executions of Android apps. Our approach (\monkeylabsp) derives feasible and fully replayable GUI-based event sequences for (un)natural app usage scenarios (i.e., actionable scenarios) based on the novel {\tt \textit{Record}}$\rightarrow${\tt \textit{Mine}}$\rightarrow${\tt \textit{Generate}}$\rightarrow${\tt \textit{Validate}} framework.  \monkeylab provides stakeholders with an automated approach for scenario generation that can be as powerful as manual testing. 
\monkeylab mines event traces and generates execution scenarios using statistical language modeling, static analysis, and dynamic analysis.
To generate (un)natural event sequences, our novel approach (i) augments the vocabulary of events mined from app usages with feasible events extracted statically from the app's source code and (ii) exploits a space of possible events and transitions with different flavors of language models capable of modeling and generating combinations of events representing natural scenarios (i.e., those observed relatively frequently in app usages) and  corner cases (i.e., those sequences that were observed relatively infrequently or not observed at all).  

We evaluated \monkeylab on several medium-to-large Android apps from Google Play and compared \monkeylab to other commonly used approaches for generating GUI-based event scenarios. The results demonstrate that \monkeylab is able to generate effective and fully replayable scenarios. Moreover, \monkeylab is able to generate scenarios that differ from observed executions enabling it to explore other paths that could trigger unexpected app crashes.

In summary, the paper provides the following contributions:
\begin{itemize}
\item A {\tt \textit{Record}}$\rightarrow${\tt \textit{Mine}}$\rightarrow${\tt \textit{Generate}}$\rightarrow${\tt \textit{Validate}} framework for generating actionable scenarios for Android apps. We designed \monkeylab to be independent from specific Android devices or API platform versions.
\item A novel mechanism for generating actionable scenarios that is rooted in mining event traces, generating execution scenarios via statistical language modeling, static and dynamic analyses, and validating these resulting scenarios using interactive executions of the app on a real device (or emulator). In particular, we explore interpolated \textit{n}-grams and back-off models, and we propose three different flavors (i.e., up, down, and strange) for generating (un)natural scenarios.
\item A thorough empirical evaluation and comparison of \monkeylab to competitive approaches on Android devices (Google Nexus 7 tablets). Experimental data, videos of the generated scenarios, and other accompanying tools are available in our online appendix~\cite{Monkeylab-OA:MSR15}. 
\end{itemize}

\section{Background and Related Work}
\label{sec:related}
Automatic generation of GUI-based scenarios (event sequences) has applications not only in automated testing but also in creating app usage documentation ~\cite{Wang:CHI14}.  In general, event sequences can be generated automatically by relying on models built (i) statically from app source code, (ii) dynamically from interactive app executions (e.g., GUI ripping) or from execution traces, (iii) manually defined by programmers, or (iv) approaches using random chains of events without any knowledge of the app such as the widely used Android \texttt{UI monkey}~\cite{Monkey}. Once the model is defined, it can be used to generate sequences of feasible (in theory) events.

Models derived from the app source code rely on static analysis \cite{Rountev:CGO14}, symbolic~\cite{Jensen:ISSTA13,Yeh:SERE-C14} and concolic execution~\cite{Anand:FSE12} techniques. Rountev \emph{et al.} \cite{Rountev:CGO14} proposed a method for   statically extracting GUI components, flows of GUI object references and their interactions. The concolic-based testing model proposed by Anand \emph{et al.}~\cite{Anand:FSE12} generates single events and event sequences by tracking input events from the origin to the point where they are handled.

Models derived using dynamic analysis are mostly based on interactive execution of the app for systematically identifying the GUI components and transitions (i.e., GUI ripping).  The execution is usually done heuristically by using some strategy (e.g., depth-first search (DFS) or a non-uniform distribution). Examples of approaches and tools relying on systematic exploration are the work by Takala \emph{et al.}~\cite{Takala:ICST11}, the tools \emph{VanarSena}~\cite{Ravindranath:MobiSys14}, \emph{AndroidRipper}~\cite{Amalfitano:ASE12},  A$^3$E~\cite{Azim:OOPSLA13},  \emph{Dynodroid}~\cite{Machiry:FSE13}, \emph{SwiftHand}~\cite{Choi:OOPSLA13},  $OME^{*}$~\cite{Nguyen:TSE14}, and \emph{MobiGUITAR} \cite{Amalfitano:IEEE14} that extracts a state-based model of the GUI, which is used to generate JUnit test cases. 

Dynamic analysis has also been used for collecting real execution traces and inferring some grammar-based rules or  statistical models describing observed events. Representative approaches of models extracted from execution traces collected apriori include  \emph{SwiftHand}~\cite{Choi:OOPSLA13}, which uses statistical models extracted from execution traces, and the approach by Elbaum \emph{et al.}~\cite{Elbaum:TSE2005} which generates test cases from user-session data. TAUTOKO~\cite{Dallmeier:TSE12} mines typestate models from test suite executions and then mutates  the test cases to cover previously unobserved behavior. Recent work by Tonella \emph{et al.}~\cite{Tonella:ICSE14} derives language models from execution traces.

Hybrid models, such as \emph{ORBIT}~\cite{Yang:FASE13}, have also been proposed, which combine static analysis and GUI ripping. Moreover, \emph{Collider}~\cite{Jensen:ISSTA13} combines GUI models and concolic execution to generate test cases that end at a target location. \emph{EvoDroid}~\cite{Mahmood:FSE14}  extracts interface and call graph models automatically, and then generates test cases using search-based techniques. Yeh \emph{et al.}  \cite{Yeh:SERE-C14} track the execution paths of randomly generated input events by using symbolic execution. Afterward, the paths are used for detecting faults and vulnerabilities in the app.

Event sequences can also be generated by relying on random distributions. For instance, the Android \texttt{UI monkey}~\cite{Monkey} is widely used for generating random input events without prior knowledge of the app's GUI. Although a tool like \texttt{monkey} helps automate testing, it is not robust since it is prone to providing a high ratio of invalid inputs. \emph{Dynodroid}~\cite{Machiry:FSE13} also includes a strategy for generating GUI and system level events via a uniform distribution where available GUI events are discovered interactively using GUI ripping.

Regardless of the underlying analyses to build the model, model-based GUI testing is inherently difficult, because the current state of an app (available components and possible actions) can change continually as new GUI events are executed. For example, consider a state machine-based GUI model in which each node (state) is a screen (i.e., activity) of an Android app, and the transitions are actions on specific GUI components (e.g., click the OK button). Some of the real transitions are feasible under certain conditions (e.g., delete a task in a non-empty list of a TODO app), and some states that are reachable only after executing natural (and unnatural) sequences of GUI events (e.g., customized scroll behavior on an empty list) are hard to model using state machines. In addition, although systematic-based exploration techniques could execute events on all of the GUI components of the app, the execution follows a predefined heuristic on the GUI (e.g., DFS) that neither represents natural execution scenarios nor considers execution history. Existing approaches also assume a high rate of generated event sequences that are simply infeasible ~\cite{Dias:WEASELTech7, Tonella:ICSE14}. For instance, some events in the sequence can be infeasible in the context of a given test case (e.g., an invalid action on a component or an action on a component that is not displayed in the current GUI state). Another issue, which is pertinent to the models extracted from execution traces, is their inability to generate sequences with unseen (i.e., not observed in the traces) but feasible events.

\section{Mining and Generating Actionable Execution Scenarios with \monkeylabsp}
\label{sec:approach}
Given that manual execution of Android apps for testing purposes is still preferred and relied upon over automated methods \cite{Mona:ESEM13}, we set out to build a system for generating inputs (event sequences) to mobile apps on Android that would simulate the convenience, naturalness, and power of manual testing (and utilize all the data that is produced during these app usages) while requiring little effort from developers. Also, we identified a set of four key goals that we felt such a system must satisfy in order to be useful:\begin{enumerate}

\item The solution for generating  executable scenarios needs to resemble manual testing, where the process of collecting test scripts allows developers/testers to interact with the app naturally by using gestures and/or GUI-based actions. One solution to this issue is represented by the event log collection capability used by \emph{RERAN} \cite{Gomez:ICSE13}. However, \emph{RERAN} replays only what it records and does not allow an arbitrary recombination of events for replay; moreover, scripts recorded with \emph{RERAN} cannot be easily deciphered by testers since they are low-level, hardware-specific events that are coupled to screen locations (e.g., events collected on a Nexus 4 will not work on a Galaxy S4).

\item The solution should have context awareness of an app's execution history, where current possible actions are generated and executed as event streams to facilitate the highest possible testing coverage. In this regard, our key insight is to rely on statistical language models (\textit{n}-grams) to generate event sequences by using program structure and previous executions of an app. Our motivation behind using language models (LMs) is in their capacity to represent event streams while maintaining contextual awareness of the event history. \monkeylab uses LMs for representing sequences of events as tokens (more specifically, \textit{Object:Action} lexemes) that can be extracted statically from the app and also mined from the app's execution. Extracting the tokens only from execution traces carries the risk of missing possible feasible events that are not observed in traces. Therefore, combining static and dynamic information may further augment the vocabulary used by LMs in such a way that some unnatural (those that did not appear in traces) but feasible tokens are included in the model, effectively taking into account both the execution history and unobserved events.

\item The solution should be able to generate \emph{actionable scenarios} as streams of events that can be reproduced automatically on a device running  the app.  These scenarios should not be low-level event streams but rather human readable streams that developers can easily comprehend and are not coupled to locations in a screen.  This goal can be met by translating high-level readable tokens (e.g., click OK button) to low-level event streams that can be executed automatically on a target device. These high-level readable tokens are derived from language models formulated by mining from app usages and source code.

\item The solution should not impose significant instrumentation overhead on a device or an app in order to be able to reproduce automatically generated scenarios. In order to meet this goal, we rely on the replay capability provided by the input commands for remote execution in Android~\cite{InputCommands}.  However, any framework for remote execution of GUI events can be used such as Robotium~\cite{Robotium}. 
\end{enumerate}	

\begin{figure*}[tb]
\begin{center}
\includegraphics[width=0.83\linewidth]{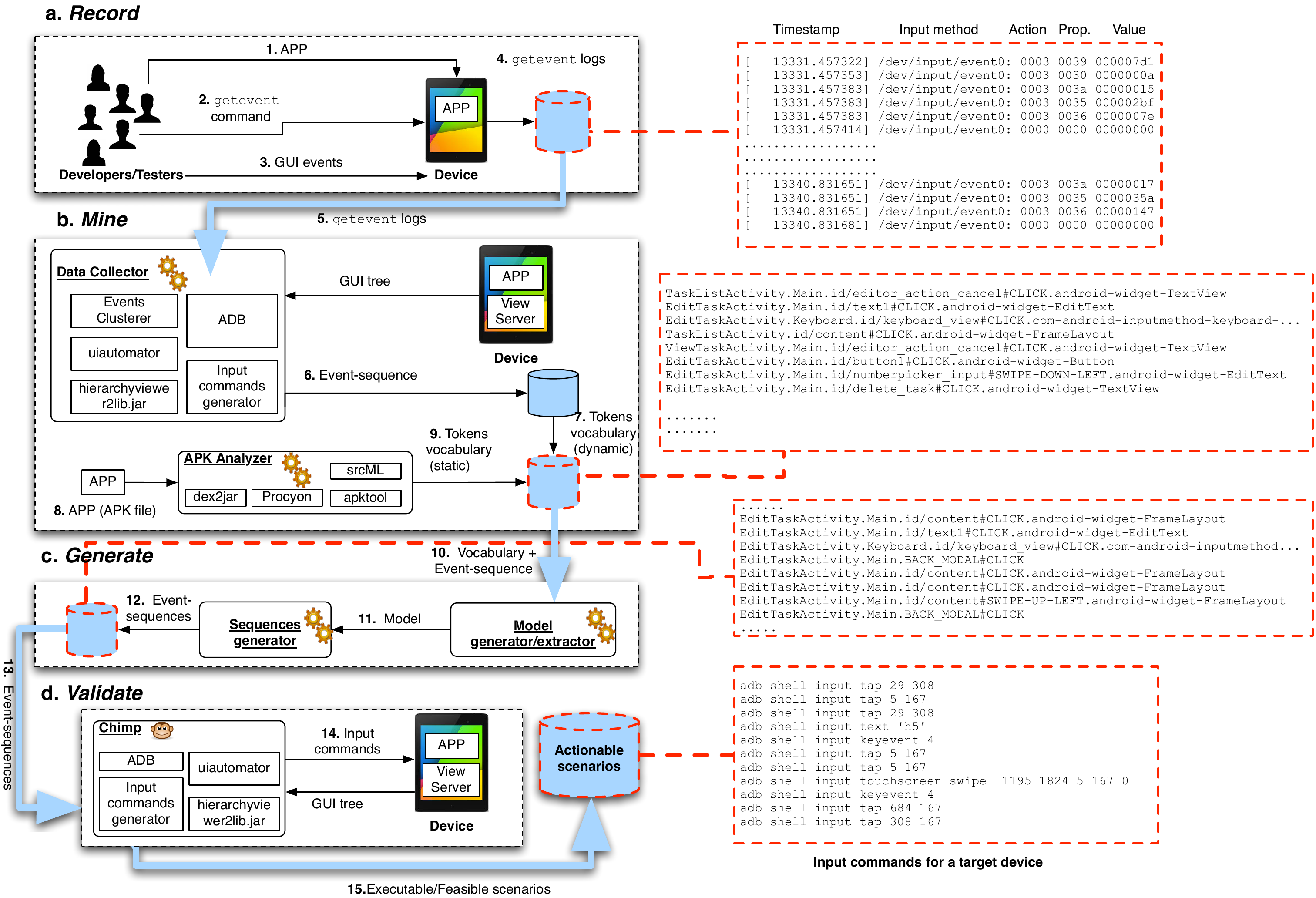}
\vspace{-1em}
\caption{MonkeyLab architecture and the {\tt \textit{Record}}$\rightarrow${\tt \textit{Mine}} $\rightarrow${\tt \textit{Generate}}$\rightarrow${\tt \textit{Validate}} framework}
\label{fig:monkeylab}
\end{center}
\vspace{-2.8em}
\end{figure*}

Having these goals in mind, the proposed approach for generating actionable scenarios is described by the following framework: (i) developers/testers use the app naturally; the event logs representing scenarios executed by the developers/testers are {\tt \textit{Recorded}}; (ii) the logs are {\tt \textit{Mined}} to obtain event sequences described at GUI level instead of low-level events; (iii) the source code of the app and the event sequences are {\tt \textit{Mined}} to build a vocabulary of feasible events; (iv) language models are derived using the vocabulary of feasible events; (v) the models are used to {\tt \textit{Generate}} event sequences; (vi) the sequences are {\tt \textit{Validated}} on the target device where infeasible events are removed for generating \emph{actionable scenarios} (i.e., feasible and fully reproducible). The framework and the architecture are depicted in Figure \ref{fig:monkeylab} as well as real examples of event logs collected in the {\tt \textit{Record}} phase, event sequences derived after the {\tt \textit{Mine}} and {\tt \textit{Generate}} phases, and actionable scenarios  generated with the {\tt \textit{Validate}} phase, for the Android app (Mileage). In the following subsections, we describe each step of the framework and the components required for each step.

\subsection{ {\tt \textit{Record}}: Collecting Event Logs from the Crowd}
Collecting event logs from developers/testers should be no different from manual testing (in terms of experience and overhead). Therefore, developers/testers using \monkeylab rely on the {\tt getevent} command for collecting event sequences, similar to \textit{RERAN} \cite{Gomez:ICSE13}. The {\tt getevent} command produces logs that include low-level events (see {\tt getevent} log example in Fig. \ref{fig:monkeylab}), representing click-based actions as well as simple (e.g., swipe) and complex gestures; those logs are collected during an app execution. After \monkeylab is enabled, developers/testers exercise/test the app as in manual testing. After having executed the app, the logs are generated with the {\tt getevent} command and copied to the logs repository. Since our log collection approach poses no overhead on developers/testers (they just use apps as usual), this setup permits collecting logs on a large scale. In fact, \textit{this log collection approach can be easily crowd-sourced, where logs are collected from daily app usages by ordinary users}. 

\subsection{{\tt \textit{Mine}}: Extracting Event Sequences from Logs and the App}
The goal of the {\tt \textit{Mine}} phase is to extract the vocabulary of events (i.e., feasible events) and translate the {\tt getevent} logs required to build the language models. The vocabulary of events is extracted from the source code of the app and also from the event logs. However, the low-level events are coupled to specific locations on the screen and do not describe the GUI components in the app; in addition, a single GUI event is represented by multiple lines in a log (see Fig.~\ref{fig:monkeylab}). A line in a {\tt getevent} log has a timestamp, input method (e.g., screen or physical keyboard), an action (e.g., {\tt 003 = click}), a property related to the action (e.g., {\tt0035} = \textit{x}-axis position), and the property value. Therefore, to eliminate the dependency of the actions on the screen coordinates in specific devices, we translated the {\tt getevent} logs to a GUI-based level representation: we model an event  $e_i$---represented by multiple lines in a \texttt{getevent} log---as the tuple \emph{$e_i: = <$Activity$_i$, Window$_i$, GUI-Component$_i$, Action$_i$, Component-Class$_i$$>$} (see examples of event sequences in Fig.~\ref{fig:monkeylab}).  We included the \emph{Window} element to distinguish actions on the activity (i.e., screen) and actions on the displayable Android keyboard. For the \emph{Action} element, we included {\tt click}, {\tt long-click}, and  {\tt swipe}. This representation considers the fact that Android apps are composed mostly of Activities representing screens of the application. Each Activity is composed of GUI components that are rendered dynamically according to the app's state. Thus, the number of GUI components that are visible in an activity can be different across time.  Additionally, the set of feasible actions for each component depends on the component type. For instance, {\tt click} and {\tt long-click} event handlers can be attached to a button but not to a {\tt swipe} gesture handler. The \emph{Component-Class} field is necessary to validate GUI-components in the app at runtime. This representation of $e_i$ tuples is used for both generating the vocabulary of events from app code and event sequences from {\tt getevent} logs.

\subsubsection{Mining GUI events statically from APKs (APK Analyzer)}
\label{sec:apka}

The \emph{APK-Analyzer} component  (Fig.~\ref{fig:monkeylab}-b) uses static analysis to extract from the app source code a list of feasible GUI events. The list is used to augment the dynamically built vocabulary from user event streams. To achieve this, GUI components are extracted from decompiled APKs before links between these components to activities, windows, and actions/gestures are constructed. To extract this information, the \emph{APK-Analyzer} (i)  uses the \textit{dex2jar}\cite{dex2jar} and  \textit{Procyon}\cite{procyon}  tools for decompilation and (ii) converts the source files to an XML-based representation using \textit{srcML} \cite{srcml}. We also rely on \textit{apktool} \cite{apktool} to extract the resource files from the app's APK.  The {\tt id}s, types, and hierarchy of the GUI components were extracted from the XML files of the APK resources.

The next major step in building the static vocabulary is linking each GUI component to its respective actions/gestures.  Rather than parsing the source code to determine which gestures are instantiated, the \emph{APK-Analyzer} assigns inherent gestures to the types of GUI components.  For standard Android component types, the linking is done with expected gestures, e.g., {\tt Button} would be linked with {\tt click} and {\tt long-click}.   For custom components, the \emph{APK-Analyzer} parses source code to determine gesture handlers and event listeners, which are attached to the custom components.  After linking actions/gestures and types to the components is complete, the \emph{APK-Analyzer} links the components to the Activities in which they appear. The list of events extracted statically from the app is represented by a set of tuples $e_i$.  It should be noted that the APK Analyzer cannot generate a static vocabulary from obfuscated code nor code where components are instantiated dynamically.

\subsubsection{Mining GUI events from event logs (Data Collector)}
\label{sec:dc}
The $e_i$ tuples require high-level information (i.e., activity, GUI component, window) that is not provided by the {\tt getevent} logs. Thus, we implemented a component (\emph{Data collector} in Fig.~\ref{fig:monkeylab}-b), which is able to translate {\tt getevent} logs into sequences of $e_i$ tokens. The \emph{Data collector} replays the logs in a ripping mode in order to dynamically collect the GUI information related to the event. 
The sentences (i.e., lines) in a log $L_k$ are grouped to identify individual GUI events. Then each group of sentences ($c$) representing a GUI event is translated into a natural language description (e.g., $<${\tt click}, $\{x=10,y=200\}$$>$). For identifying the corresponding component in the GUI, we queried the {\tt Android View Server} running on the device. The {\tt View Server} provides a tree representation of the GUI components with some attributes such as the top-left location of a component and the dimensions. Given the event location, we traversed the GUI tree looking for the component area (top-left corner plus dimensions) that contains the event's location. For identifying the current activity we used the {\tt adb shell dumpsys window windows} command, and then we looked for the property {\tt mCurrentFocus|mFocusedApp}. For identifying whether the keyboard is displayed, we queried the list of current windows in the {\tt View Server}. With the $e_i$ tuple, we built an input event command\cite{InputCommands} that can be remotely executed using the {\tt Android Debugger Bridge} (e.g., {\tt adb shell input tap x y}). Finally, we executed the input command to update the app state and continued with the next group $c$. The output of the procedure is the list $T_k$ of $e_i$ tuples extracted from the  $L_k$ {\tt getevent} log. The $T_k$ sequences and unique $e_i$ tuples are copied to the sequence and token repositories, respectively.
 
\subsection{{\tt \textit{Generate}}: Event Sequences with Language Models}
The vocabulary of event tokens extracted from the app code and the event logs, both represented as GUI-level events, are the universe of feasible events in the app and are natural use cases respectively.  They represent the GUI model of the app including activities, components, feasible actions, and natural transitions. This data can be used to build statistical models for generating streams of events as execution scenarios of the app. In fact, the {\tt \textit{Generate}} phase of our framework uses LMs \cite{Jelinek:1980, Katz:1987,Church:CSL91} to generate (un)natural sequences of events for that purpose. These LMs are trained with the vocabulary and event sequences extracted in the {\tt {\textit Mine}} phase.  

Our motivation behind using LMs is their capacity to represent event streams while maintaining a contextual awareness of the event history. Hindle \textit{et al.} \cite{Hindle:ICSE12} demonstrated the usefulness of applying LMs to software corpora because real programs written by real people can be characterized as \emph{natural} in the sense that they are even more repetitive than natural languages (\textit{e.g.}, English). Tonella \emph{et al.} \cite{Tonella:ICSE14} exploited this interpretation of naturalness by casting it to streaming events as test cases. \textit{Our hypothesis is that event-sequences extracted from {\tt getevent} logs emit a natural set of event streams, which map to core functionality, as well as an unnatural set of event streams}. Language models inherently capture natural behavior, but we also consider the complement of this natural space, thereby improving the interpretive power of these statistical models. Our novel interpretation of software LMs systematically segments the domain of a LM over dynamic traces and imputes testing semantics to each segment. For example, a sampling engine may select tokens from the natural space to improve coverage, yet it may sample from the unnatural space to crash the app. In the following subsections we provide formal definitions behind LMs, LM flavors that we defined in the context of our work as well as details on using LMs for generating event sequences.

\subsubsection{Language Models}
\label{sec:lm}
A statistical language model is a probability distribution over units of written/spoken language,which measures the
probability of a sentence $s=w_1^m=w_{1}w_{2}...w_{m}$ based on the words (a.k.a., tokens) probabilities:
\begin{equation}
\label{eq:LanguageModel}
p(s) = p(w_1^m) = \prod_{i=1}^m p(w_i|w_1^{i-1}) \approx \prod_{i=1}^m p(w_i|w_{i-n+1}^{i-1})
\vspace{-0.3em}
\end{equation}
In Eq.~\eqref{eq:LanguageModel}, the Markov assumption approximates the joint distribution by measuring the conditional probabilities of token subsequences known as $n$-grams. After ``discounting'' the model's maximum likelihood estimates, there are generally two methods for distributing the probability mass gleaned from the observed $n$-grams: back-off and interpolation. The Katz back-off model \cite{Katz:1987} takes the form
\begin{equation}
p_{\mathcal{B}}(w_i|h) =
\begin{dcases}
\alpha(w_i|h) & c(hw_i) > k
\\
\beta(h) p_\mathcal{B}(w_i|w_{i-n+2}^{i-1}) & \text{otherwise}
\end{dcases}
\vspace{-0.2cm}
\end{equation}
where $h = w_{i-n+1}^{i-1}$ is the history, $\alpha(w_i|h)$ is the discounted maximum likelihood estimate of word $w_i$ given $h$, $\beta(h)$ is the back-off weight, $c(hw_i)$ is the number of times $hw_i$ appears in the training corpus, and $k$ is a constant, which is typically zero. If the history was observed in training, then a back-off model says the conditional probability of a word given its history is equal to the discounted estimate of the $n$-gram, where $\alpha(w_i|h)$ is computed using a smoothing technique such as Good-Turing estimation \cite{Good:1953}. Otherwise, the model truncates the history and recursively computes the probability. Back-off models only consider lower-order $n$-grams when $c(hw_i) = 0$ for every $w_i$ in the vocabulary $\mathcal{V}$. On the other hand, interpolation considers lower-order $n$-grams whether or not $c(hw_i) > 0$. The general interpolated model takes the form
\begin{equation}
\label{eq:InterpolatedModel}
p_{\mathcal{I}}(w_i|h) = \alpha(w_i|h) + \beta(h) p_\mathcal{I}(w_i|w_{i-n+2}^{i-1})
\end{equation}
where $\alpha(w_i|h)$ is computed using a smoothing technique such as modified Kneser-Ney estimation \cite{Chen:1996}. 

\subsubsection{Language Model Flavors}
\label{sec:brain_lm}
Back-off ({\tt BO}) and interpolation ({\tt INTERP}) are different approaches for computing probabilities, but each LM is a way to characterize the naturalness of a sentence. Moreover, we propose that each model can be used to generate unnatural sentences as well, and these unnatural sentences have clear software testing semantics. Parenthetically, both {\tt BO} and {\tt INTERP} can be decomposed to different flavors---{\tt up}, {\tt down}, and {\tt strange}---that are simple transformations for driving a sampling engine to specific segments of the language model's domain.

{\tt up} corresponds to the natural distribution over tokens, but there are subtle differences in the implementation depending on whether the language model uses {\tt BO} or {\tt INTERP}. For example, suppose we are given the following conditional probabilities from a LM: $\theta = \{ \alpha(w_a|h) : 0.20, \alpha(w_b|h) : 0.50, \alpha(w_c|h) : 0.10, \alpha(w_d|h) : 0.10 \}$. For {\tt BO} models, {\tt up} will sample a uniform variate using the cumulative sum of the smoothed estimates. If $c(hw) = 0$, then {\tt up} will back-off to the next lowest order and repeat the procedure, where unigram probabilities serve as the base case in the recursion. By construction, interpolation mixes input from every $n$-gram that can be sliced from $hw$, so {\tt up} recurses on the order through the unigrams to compute $p(w_i|h) \forall w_i \in \mathcal{V}$ using Eq.~\eqref{eq:InterpolatedModel}. Naturally, while the probability mass is concentrated on expected transitions, {\tt INTERP-up} models are able to reach any transition in the vocabulary of transitions at any point in time. This is not the case for {\tt BO-up} models.

{\tt down} corresponds to the unnatural distribution. For example, suppose we are given the same conditional estimates $\theta$. For {\tt BO} models, {\tt down} will normalize the estimates, sort the tokens according to their probabilities and then reverse the probabilities over the tokens to produce the following distribution: $\{ p(w_c|h) : 0.55, p(w_d|h) : 0.22, p(w_a|h) : 0.11, p(w_b|h) : 0.11 \}$. In cases where two tokens have the same probability, {\tt down} will randomly arrange the tokens, so $\{ p(w_c|h) : 0.55, p(w_d|h) : 0.22 \}$ and $\{ p(w_c|h) : 0.22, p(w_d|h) : 0.55 \}$ are equally likely. According to the training log, tokens $w_c$ and $w_d$ are unnatural given the context, but {\tt down} seeks the unnatural transitions when it generates samples. For {\tt INTERP} models, {\tt down} will build a categorical distribution over $\mathcal{V}$ using Eq.~\eqref{eq:InterpolatedModel}. Therefore, as with {\tt up}, we sort the tokens in $\mathcal{V}$ according to their probabilities and then reverse the probabilities over the sample space. {\tt INTERP} {\tt down} models marshal uncharacteristic transitions in $\mathcal{V}$ given the context into the natural space, replacing frequently observed transitions.

{\tt strange} linearly interpolates {\tt up} and {\tt down}, which likely has the effect of increasing the entropy of $p(w|h)$. When the entropy of the natural distribution over tokens is low and the mixture coefficient $\lambda \approx 0.5$, {\tt strange} distributions can take an interesting form where the probability mass is simultaneously concentrated on very (un)natural events. This strange case describes a scenario where the generative model produces a stream of natural tokens, yet at any point in time is "equally" likely to choose a corner case in the execution.

\subsubsection{Generating Event Sequences}
\label{sec:brain_tc}

The generation engine estimates two $n$-gram language models---one {\tt BO} and one {\tt INTERP}--- using the vocabulary extracted in the \emph{Mine} phase. 
The default order and smoother for both models are three and modified Kneser-Ney \cite{Chen:1996}, respectively. Tokens in the vocabulary that are not in the event-sequence repository are factored into the models' unigram estimates. Clients (i.e., instances of the \emph{Chimp} component in Fig.~\ref{fig:monkeylab}) can communicate with the \emph{Sequence Generator} by sending JSON requests and reading JSON responses. A request includes: model, flavor, history, and length. The model field can take one of two values: {\tt BO} or {\tt INTERP}. The flavor field can take one of three values: {\tt up}, {\tt down} or {\tt strange}. The history field contains the initial prefix as an array of strings. If the client does not have a prefix, then the history field is the empty string. When the client does not have a history (or the length of the history is suboptimal), then the \emph{Sequence Generator} will essentially bootstrap the history to length $n-1$, where $n$ is the order of the model(s). When the client gives a history that is longer than $n$, the \emph{Sequence Generator} will apply the Markov assumption, so the history will be sliced according to the order before querying the language model. The length field governs the length of the sequence to query from the language model (i.e., number of events in the test case). This enables us to use the \emph{Sequence Generator} in serial mode (e.g., requesting a sequence with 100 events) or in interactive mode (i.e., only one event is requested at a time).

\subsection{{\tt \textit{Validate}}: Filtering Actionable Scenarios}
The streams of events generated by the \emph{Sequence Generator} are expressed at GUI level, which decouples the sequence from device-specific locations.  Thus, the \emph{Chimp} component (Fig.~\ref{fig:monkeylab}-d) understands the events and {\tt \textit{Validates}} them dynamically on a device. The validation can be performed in a serial (as in Tonella et al. \cite{Tonella:ICSE14}) or with \monkeylab interactive mode.

\begin{algorithm}[t]
\label{alg:chimp-serial}
\LinesNumbered 
\KwIn{$S$}
\KwOut{$AS$}
\Begin{
	$i = 1, ATC = \emptyset$\;
	$startAppInDevice()$\;
	\ForEach{$e \in S$} {
		$feasible = queryVS(e.component,e.action)$\;
		\If{!feasible}{
			$continue\_with\_next\_event$\;
		}
		
		$<x,y>=searchVS(e.component)$\;
		$cmd_i =getInputCmd(e.action, <x,y>)$\; 
		$addEvent(AS, <e,cmd_i>)$\;
		$executeInputCmd(cmd_i)$\;
		$i = i + 1$\;
	}
}
\caption{Validating Event Sequences: Serial Mode}
\end{algorithm}
In the serial mode, the \emph{Chimp} requests a sequence that is validated iteratively following the procedure in  Alg.~\ref{alg:chimp-serial}.  For each event $e$ in the event sequence $S$, the \emph{Chimp} searches for the component (in the event tuple) in the {\tt Hierarchy View} (i.e., GUI tree) of the current GUI displayed in the target device. If the component is not in the GUI or the current activity is not the same as in the tuple $e$ (line 5), then the \emph{Chimp} discards the event (line 7); otherwise the \emph{Chimp} queries the {\tt View Server} (line 8) in the device to identify the location of the component. Afterward, it generates an {\tt input} command by using the location $<$$x,y$$>$ of the component and the action in the tuple $e$ (line 9). Then the event $e$ and its corresponding input command are added to the actionable scenario (line 10).  Finally, the command $cmd_i$ is executed on the GUI to update the app state. This serial model is similar to the one proposed in \cite{Tonella:ICSE14}, however we are proposing three different flavors for the interpolated \textit{n}-grams and the back-off model (Sec.~\ref{sec:lm} and Sec.~\ref{sec:brain_lm}) and we automatically validate the sequences on a target device.

\begin{algorithm}[t]
\label{alg:chimp-interactive}
\LinesNumbered 
\KwIn{$k$}
\KwOut{$AS$}
\Begin{
	$ATC = \emptyset, history=\emptyset$\;
	$startAppInDevice()$\;
	\For{$i \in 1:k$} {
	        $e=queryEvent(history)$

		$feasible = queryVS(e.component,e.action)$\;
		
		\If{$!feasible$}{
			
			$e = getEvent(randComponent())$\;
			
		}

		$<x,y>=searchVS(e.component)$\;
		
		$cmd_i =getInputCmd(e.action, <x,y>)$\; 
		
		$addEvent(AS, <e,cmd_i>)$\;
		$history = e$\;	
		$executeInputCmd(cmd_i)$\;
		$i = i + 1$\;
	}
}
\caption{Validating Sequences: Interactive Mode}
\end{algorithm}

In the interactive mode (Alg.~\ref{alg:chimp-interactive}), the \emph{Chimp} requests single events until the target length $k$ is exhausted. Our motivation for this mode is the possibility of having infeasible events in a sequence (i.e., in serial model) that can influence further events. We are augmenting the vocabulary used to train the models, with individual tokens extracted statically from the source code. So, it is possible that unseen---but also infeasible---events appear in a sequence.  For example, in the sequence $e_1e_2...t_1e_ke_{k+1}...t_2t_3e_le_{l+1}$, tokens $t_1$, $t_2$, and $t_3$ are infeasible, which leads to the following issues: (i) the inclusion of events $e_k$, $e_{k+1}$, $e_l$, $e_{l+1}$ into the sequence is influenced by the infeasible tokens $t_1$, $t_2$, and $t_3$; and (ii) once the first infeasible token is read by the \emph{Chimp}, the further events in the sequence can also be infeasible in the GUI.  This is why, in the serial mode, we opted to skip infeasible events and continue reading until the sequence is exhausted, which drives to sequences with less events than the target $k$. 

Therefore, as a second option, we opted for requesting single events and executing only feasible events. This is how the interactive mode operates. The \emph{Chimp} in interactive mode asks for a single token (similarly to code suggestion problem \cite{Hindle:ICSE12}) until a target number of feasible events (lines 4 and 5 in Alg.~\ref{alg:chimp-interactive}) is executed on the device. To avoid infeasible events and loops because of chains of infeasible events (i.e., the returned token is always infeasible), we execute a random feasible event---queried from the current GUI state--- (line 7) when the component and action in event $e$ are not feasible (line 6).  This guarantees that the \emph{Chimp} always executes a feasible event, and that event is the history for the next one (line 11). The interactive mode relies on the language model extracted from the event logs and the source code, and takes advantage of the current GUI state to reduce the rate of infeasible-events produced by the model. In summary, our interactive mode combines static and dynamic analyses as well as GUI ripping.

\section{Empirical Study Design}
\label{sec:study}
Our main hypothesis is that models derived with \monkeylab  are able to produce event sequences for natural scenarios and corner cases unseen in the scenarios executed by stakeholders. Therefore, event sequences generated with \monkeylab should not only include events from scenarios used to derive the models (i.e., scenarios collected from users) but also events not observed in those scenarios. To test our hypothesis, we performed a case study with five free Android apps and unlocked/rooted Nexus 7 \cite{Nexus7} Axus tablets each having a 1.5GHz Qualcomm Snapdragon S4 Pro CPU and equipped with Android 4.4.2 (kernel version 3.4.0-gac9222c). 

We measured statement coverage of sequences generated with (i) \monkeylabsp, (ii)  a random-based approach (i.e.,  {\tt Android GUI monkey}),  (iii) GUI ripping using a DFS heuristic, and (iv) manual execution.  We also compared the GUI events in the sequences to identify events that were executed only by one approach when compared to another (e.g., GUI events in sequences from MonkeyLab but not in the users' traces). In summary, the study aimed at investigating the following research questions (RQs):
\begin{myquote}
\noindent \textbf{RQ$_1$}: \emph{Which language model-based strategy  is more suitable for generating effective (un)natural scenarios?}\\
\noindent \textbf{RQ$_2$}: \emph{Do scenarios generated with \monkeylab achieve higher coverage as compared to {\tt Android UI monkey}?} \\
\noindent \textbf{RQ$_3$}: \emph{How do scenarios generated with \monkeylab compare to a DFS-based GUI ripper in terms of coverage?}\\
 \textbf{RQ$_4$}: \emph{Do  \monkeylab scenarios achieve higher coverage than manual executions used to train the models?}
\end{myquote}

\subsection{Data Collection}
\label{sec:data_extraction}
Tab.~\ref{tab:apps1} lists the Android apps used in the study, lines of code (excluding third party libraries), number of Activities,  methods, and GUI components. We selected the applications looking for a diverse set in terms of events supported by the apps ({\tt click}, {\tt long click}, {\tt swipe}), number of GUI components, number of Activities, and application domain. However, the main selection criteria was the app's size. We decided to use only medium-to-large applications that exhibit non-trivial use-cases and a large set of feasible events.

To answer the research questions, we simulated a system-level testing process where we asked graduate students to get familiarized with the apps for five minutes (exploratory testing), and then to test the app executing multiple scenarios for 15 minutes (functional testing).  Five Ph.D. students at the College of William and Mary executed the {\tt {\textit Record}} phase of our \monkeylab framework.   Afterward, we executed the {\tt \textit{Mine}}$\rightarrow${\tt \textit{Generate}}$\rightarrow ${\tt \textit{Validate}} phases in \monkeylab to generate actionable sequences for each app.  During the  \textit{Validate} phase, we collected coverage measurements for each of the generated event sequences by using a tailored version of Emma~\cite{Emma} for Android. Tab.~\ref{tab:apps1} lists the total number of low-level and GUI-level events collected from the participants.

To represent a random approach, we generated sequences of touch events with {\tt Android GUI monkey}, simulating  test cases composed of random events.  Regarding the GUI ripping approach, we implemented our own version of DFS-based exploration. In terms of GUI-model extraction/inference, it considers cases that were not captured in other tools \cite{Machiry:FSE13,Azim:OOPSLA13}, such as pop-up windows in menus/internal windows, on-screen keyboard,  and containers. In this case, we do not have a set of actionable sequences because the app is explored trying to visit as many clickable GUI components as posible. However, we measured the accumulated coverage for the whole systematic exploration. This measurement helped us to establish a baseline to validate whether our event sequence generation approach outperforms competitive approaches. More details of the DFS implementation are in our online appendix.
 
\begin{table}[tb]
\centering
\small
\caption{Android Apps Used in our Study. The stats include the number of activities, methods, and GUI components. Last two columns list the number of raw events (\#RE) in the event logs (i.e., lines in the files collected with the \texttt{getevent} command) and GUI level events mined from the raw logs (\#GE)}
\vspace{-0.6em}
\label{tab:apps1}
\setlength{\tabcolsep}{0.3em}
\begin{tabular}{lrrrrr|rr} \hline
App &Ver.&LOC&\#Act.&\#M.&\#Comp.&\#RE&\#GE\\ \hline
\emph{Car Report}&2.9.1&7K+&6&764&142&23.4K+&1.5K+\\
\emph{GnuCash}&1.5.3&10K+&6&1,027&275&14.7K+&895\\
\emph{Mileage}&3.1.1&10K+&51&1,139&99&9.8K+&783\\
\emph{My Expenses}&2.4.0&24K+&17&1778&693&20.3K+&854\\
\emph{Tasks}&1.0.12&10K+&4&561&200&70.6K+&1.7K+\\
\hline
\end{tabular}
\end{table}

\subsection{Design Space}
\label{sec:ds}
For training the language models, we used the traces collected from the participants for the functional testing scenario. Given  two LMs ({\textit{i.e.}, {\tt INTERP} and {\tt BO}) and three flavors (\textit{i.e.}, {\tt up}, {\tt down}, {\tt strange}),  we generated 100 scenarios for each combination $<$\emph{Model, Flavor, APP}$>$ with \textit{3}-gram LMs, then we generated actionable scenarios using the serial mode. For the interactive mode, 
we generated 33 scenarios  for each of the three INTERP flavors.  Each scenario  (in serial and interactice modes) was composed of 100 events. We also executed  {\tt Android UI monkey} on each app 100 times with 100 touchable events and inter-arrival delay of 500ms. Finally, we executed our DFS-based GUI ripper on each app.

\section{Results and Discussion}
\label{sec:results}
Fig.~\ref{fig:gnucash_cov} shows cumulative coverage of \monkeylab and Android \texttt{UI Monkey} for the event sequences, in addition to coverage achieved by logs collected from human participants and coverage of the DFS-based exploration for the \emph{GnuCash app}. The figures for the other apps are in our online appendix. In the case of \monkeylabsp, the figure depicts the accumulated coverage of the best LM in serial mode (in red color), and the combination of the 99 scenarios of \texttt{INTER-up}, \texttt{INTER-down}, and \texttt{INTER-strange} (33 scenarios for each strategy) in interactive mode (\emph{I-LM}). In the following subsections, we present the results for the research questions defined in Sec.~\ref{sec:study}.

\begin{table}[tb]
\centering
\small
\caption{Accumulated Statement Coverage of the LMs}
\vspace{-0.6em}
\label{tab:apps}
  \setlength{\tabcolsep}{0.3em}
\begin{tabular}{lrrrrrrr} 
\hline
&\multicolumn{3}{c}{\tt INTERP}&\multicolumn{3}{c}{\tt BO}&\emph{I-LM}\\
App &{\tt up}&{\tt down}&{\tt str.}&{\tt up}&{\tt down}&{\tt str.}&\\ 
\hline
\emph{Car Report}&12\%&14\%&13\%&11\%&11\%&11\%&\textbf{30\%}\\
\emph{GnuCash}&7\%&6\%&7\%&7\%&7\%&7\%&\textbf{22\%}\\
\emph{Mileage}&\textbf{36\%}&10\%&25\%&32\%&24\%&24\%&26\%\\
\emph{My Expenses}&10\%&10\%&10\%&10\%&10\%&10\%&\textbf{26\%}\\
\emph{Tasks}&40\%&34\%&35\%&22\%&22\%&34\%&\textbf{42\%}\\
\hline
\end{tabular}
\end{table}

\begin{figure}[t]
\begin{center}
\includegraphics[width=\columnwidth]{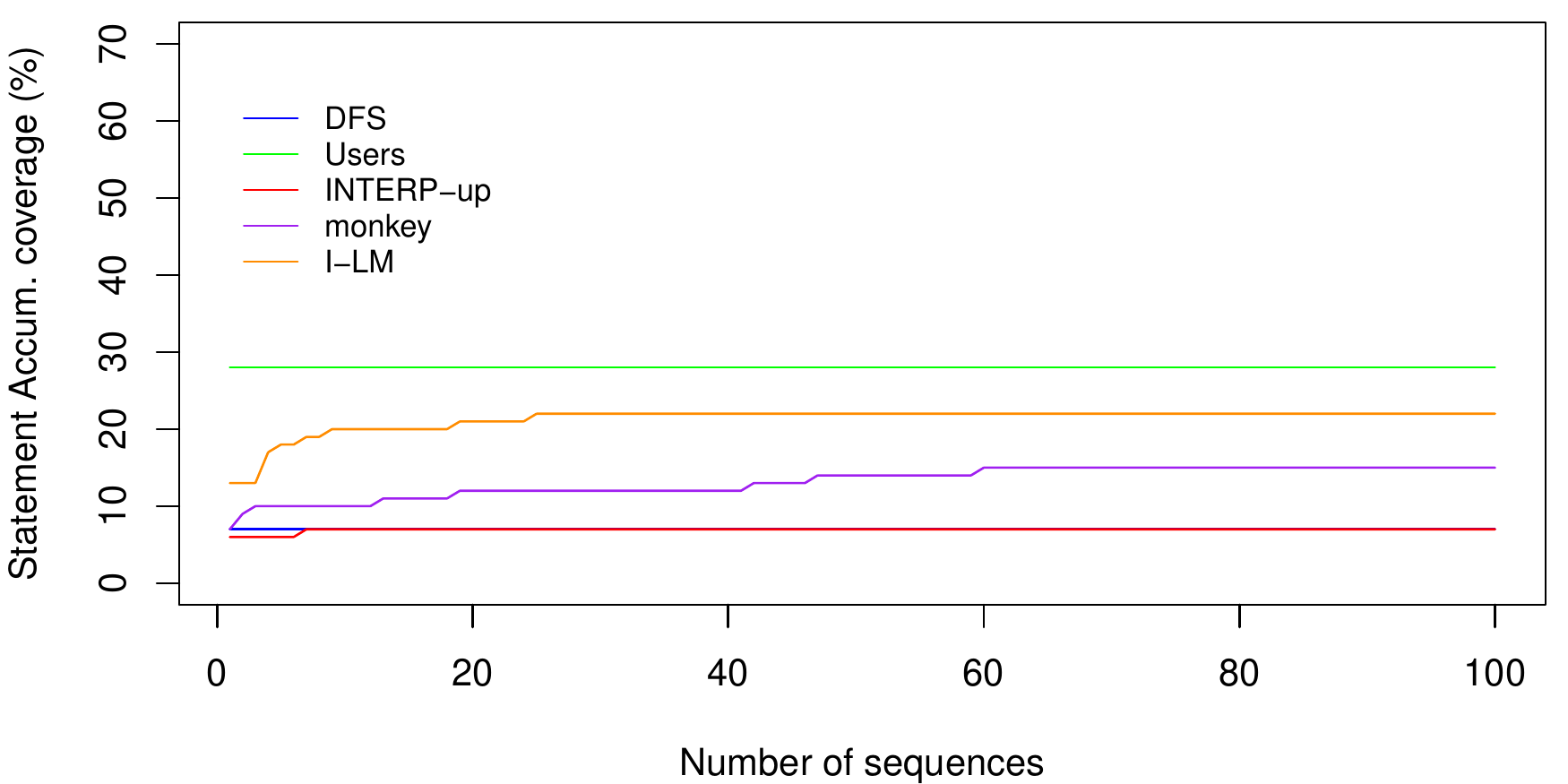}
\vspace{-0.6cm}
\caption{Accumulated Coverage for GnuCash}
\label{fig:gnucash_cov}
\vspace{-0.2cm}
\end{center}
\end{figure}

\subsection{\textbf{RQ$_1$}: Language Models and Flavors}
The accumulated coverage values for the language models in serial mode are listed in Tab.~\ref{tab:apps}. In general, the \texttt{INTERP-up} flavor is the one with the highest accumulated coverage for the serial mode. Apps with splash-screens and modal dialogs in clean launches (i.e., the app is launched right after being installed) such as \textit{GnuCash}, \textit{Car Report}, and \textit{My Expenses} posed a challenge for the language models (see Sec.~\ref{sec:lessons} for more details}). The interactive mode (\emph{I-LM}), outperforms the language models in serial model in 4 out of the 5 apps.  The \emph{I-LM} mode is able to deal with the problem of splash-screens and modal dialogs in clean launches to reduce the notoriously high rate of infeasible events.

\begin{figure}[t]
\begin{center}
\includegraphics[width=0.8\columnwidth]{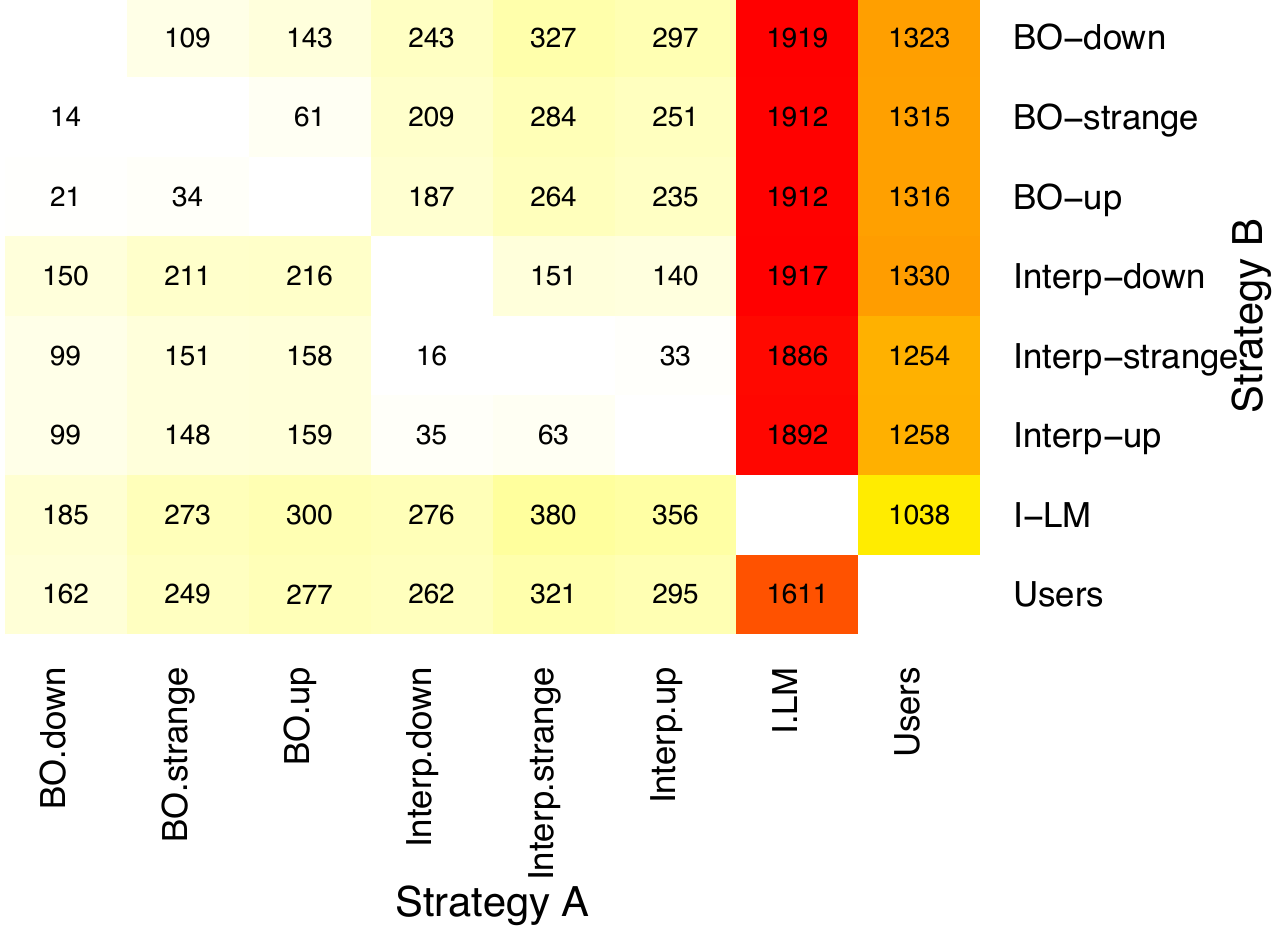}
\vspace{-0.5cm}
\caption{Total number of events executed by Strategy A that are not executed by Strategy B. The key color goes from white (zero) to red (highest value)}
\label{fig:heatmap-events}
\vspace{-0.2cm}
\end{center}
\end{figure}

\begin{figure}[t]
\begin{center}
\includegraphics[width=0.7\columnwidth]{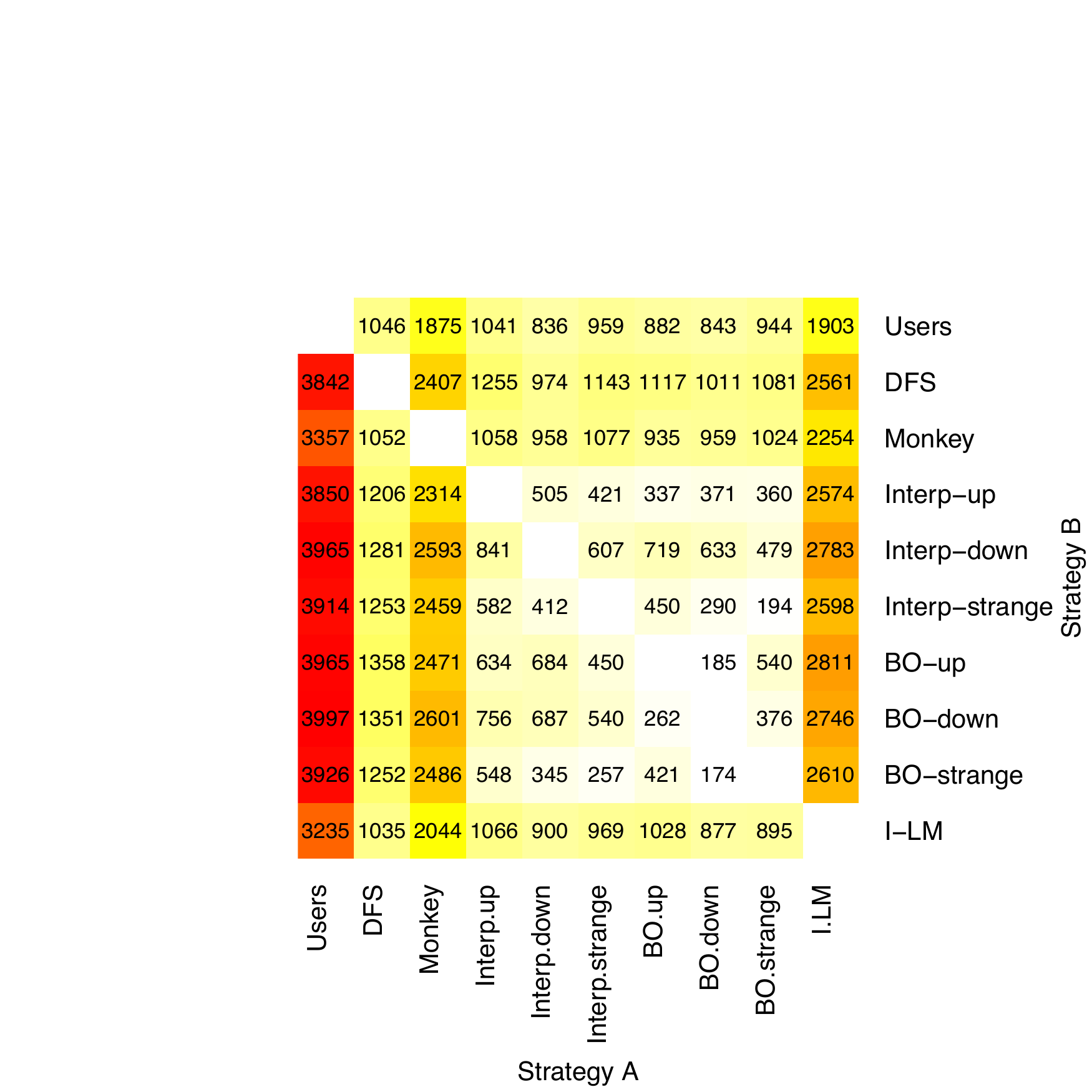}
\vspace{-0.5cm}
\caption{Total number of source code methods in which coverage is higher when comparing coverage of Strategy A versus Strategy B}
\label{fig:heatmap-methods}
\vspace{-0.2cm}
\end{center}
\end{figure}

The accumulated coverage provides a high-level measure when comparing two different models, however, it  does not reveal the entire story. It is possible that different strategies are generating different events and sequences that drive to different executions of the apps. In fact, our goal behind proposing six different flavors of LMs and the interactive mode is based on the possibility that each flavor can explore different regions in the GUI event space. Therefore, we investigated the mutually exclusive events executed by each strategy in the five apps. We also measured the number of times the method coverage for each method in an app was higher in one strategy when doing pairwise comparisons (e.g., \texttt{INTER-up} versus \texttt{INTER-down}). The results are presented with heat-maps in Fig.~\ref{fig:heatmap-events} and Fig.~\ref{fig:heatmap-methods}. Both figures corroborate the fact that each LM strategy is able to generate different sets of events. For example, when comparing {\tt INTERP-strange} to {\tt BO-down} (Fig.~\ref{fig:heatmap-events}), the former strategy was able to generate 327 GUI events that were not generated by the latter. In addition, the \emph{I-LM} mode is the strategy with the highest difference of executed events when compared to the other strategies.

\noindent \textbf{Summary for RQ$_1$:} The LMs strategies are able to generate diverse and orthogonal sets of GUI events, and the interactive mode is the most effective.

\subsection{\textbf{RQ$_2$}: \monkeylab vs. Android {\tt UI monkey}}

The coverage of \texttt{UI monkey} surprisingly outperformed the LMs in serial model for \textit{Gnu Cash}, \textit{Car Report}, and \textit{MyExpenses}.  LMs provided a better coverage only in \textit{Tasks} and showed similar coverage in \textit{Mileage}. The random nature of the events generated by the \texttt{UI monkey} allows this strategy to execute more GUI components without deeply exploring execution paths that are related to the use cases. For instance,  \texttt{UI Monkey} is able to click on a diverse set of GUI components; however, it does so without context awareness of the execution history. The LMs are able to explore execution paths that lead to the activation of  Activities and features that can not be reached easily by the \texttt{UI Monkey}. However, the \emph{I-LM} mode outperforms the accumulated coverage of \texttt{UI monkey} in all the apps except for \textit{Mileage.} When looking into the results in Fig.~\ref{fig:heatmap-methods}, it is clear how \texttt{UI monkey} is able to achieve higher coverage in source code methods more times than the serial LM-strategies. However, the \emph{I-LM} is able to achieve higher coverage in more source code methods when compared to \texttt{Android UI monkey}.

\noindent \textbf{Summary for RQ$_2$:} \texttt{Android UI monkey} is able to achieve higher coverage than \monkeylab in serial model. However, our interactive mode outperforms \texttt{UI monkey}. In terms of scenarios, all the \monkeylab strategies are able to generate execution paths that are not covered by \texttt{UI monkey}.

\subsection{\textbf{RQ$_3$}: \monkeylab vs. DFS}
\vspace{0.2cm}

The LMs in a serial mode provided similar statement coverage as compared to DFS in \texttt{GnuCash}, \texttt{Tasks}, and \texttt{MyExpenses}, better coverage in \texttt{Mileage}, and lower in \texttt{Car Report}. DFS follows a systematic strategy for executing click events on the GUIs. This systematic exploration is able to exercise deeper executions paths as compared to \texttt{UI monkey}, but DFS is not able to recognize data dependencies. We found that the LMs can actually model some of these dependencies (e.g., some fields are required to create a budget or task). In other cases, the LMs can get stuck in specific paths that are very common (frequent) in the observed traces. However, the \emph{I-LM} mode outperformed the accumulated coverage of DFS in the 5 apps. Concerning, the source code methods with higher coverage (Fig.~\ref{fig:heatmap-methods}), there is a notorious difference between DFS and the LMs in serial mode. However, the \emph{I-LM} provides higher coverage on more methods when compared to DFS (\emph{I-LM}$-$DFS = 2,561 and DFS$-$\emph{I-LM} = 1,035 methods).

\noindent \textbf{Summary for RQ$_3$:} Although the LMs in serial mode are not able to achieve better coverage than DFS, \emph{I-LM} is able to achieve higher coverage as compared to DFS.
\subsection{\textbf{RQ$_4$}: \monkeylab vs. Manual execution}
The coverage achieved by the users during the \texttt{Record} phase outperforms \texttt{UI Monkey}, DFS, and the LMs in serial and interactive modes, as originally expected. However, the LMs were able to execute events  that were not observed in the event traces collected by the users (see Fig. \ref{fig:heatmap-events}).  In general, the LMs were able to generate actionable scenarios for natural executions, i.e., there are sequences generated with the LMs including GUI events from the collected traces. However, the LMs were also able to generate sequences with unseen events, which means that the actionable scenarios cover natural cases but also corner-cases not considered by the users. And the benefit is noticeable when using the interactive mode, \textit{which was able to generate 1,611 GUI events not considered by the participants in our study.}

While we can not claim that the LMs are better than the other approaches in terms of coverage, after looking into the details of the events executed by each method, it becomes clear that the combinations of manual testing and automated approaches can significantly improve the coverage. However,  \monkeylab is able to learn a model and then generate actionable scenarios that can be executed on a target device. In addition, \monkeylab generates not only natural scenarios but also corner cases that are not generated by any of the competitive approaches.

\noindent \textbf{Summary for RQ$_4$}: Although the overall coverage achieved by \monkeylab is not as high as compared to manual execution, \monkeylab is able to generate actionable scenarios including not only natural GUI events, but also events that are not considered during manual testing. Therefore, the scenarios generated by \monkeylab can help increase the coverage achieved by manual testing without extra effort that is imposed by collecting test-scripts.

\subsection{Limitations}
\label{sec:lessons}
Automated GUI-based testing approaches can benefit from the following information in this study, since some of these issues can pose similar problems for various testing \emph{strategies}.

\textbf{Encapsulated components.}~Encapsulated components (e.g., {\tt KeyboardView}, {\tt AutoCompleteTextView}, {\tt CalendarView}, and {\tt DatePicker}) cannot be analyzed during systematic exploration, because the sub-components (e.g., each key of the keyboard) are not available for ripping at execution time. Thus, the sub-components are not recognized by the {\tt View Server} nor the {\tt Hierarchy Viewer}. This type of component requires a predefined model of the feasible events with the locations and areas of these sub-components.

\textbf{Training corpus size in serial mode.}~Another lesson gleaned from this study concerns an effective training set. 
Specifically, we diagnosed two critical issues that contributed to the high rate of infeasible events produced by the interpolated $n$-grams and back-off models.
The principal issue was relatively small amount of training data for each app as compared to the size of the respective vocabularies.
The need for substantial training data was observed in most of the sequences for the apps (especially \emph{Keepscore}) for which the LM-based strategies were not able to produce a response.
For example, an app with a vocabulary size of 250 events would require 250$\times$249 and 250$\times$250$\times$249 parameters to reliably estimate bigram and trigram models, respectively.
Data sparsity is a key concern in statistical language modeling, and a number of techniques have been developed to manage the problem of deriving useful statistical estimates from relatively small corpora~\cite{Chen:1996}.
Naturally, one approach to controlling the number of parameters is to reduce the order of the model~\cite{Manning:1999}; however, reducing the order of our models presented another issue.
In our experiments, reducing the order degenerated the effectiveness of the \texttt{INTERP-up} models at generating feasible test cases.
This degeneration can be attributed to allocating too much weight to the unigrams in the mixture model (Eq.~\ref{eq:InterpolatedModel}).
In other words, after unfolding the recurrence in Eq.~\ref{eq:InterpolatedModel}, unigrams will generally have more influence on the predictions of interpolated $n$-gram models as compared to interpolated models whose order is greater than $n$.
For example, unigrams will have more influence on the predictions of interpolated trigrams than interpolated 4-grams, because the unigrams are multiplied by $\beta(h)$ (Eq.~\ref{eq:InterpolatedModel}), a number less than one, an additional time in interpolated 4-grams.
Our expectation was for \texttt{INTERP-up} models to yield favorable coverage results in line with previous empirical studies~\cite{Tonella:ICSE14}, yet the low-orders disabled the models in our experiments.
However, we did not expect \texttt{INTERP-down} nor \texttt{INTERP-strange} models to yield high rates of feasible events.
Considering the nature of the interpolation smoothing technique, models at \emph{all} orders are factored into the estimator.
So, regardless of the order of the interpolated model and the history of events, every event will have a chance to be selected at each point in time.
Moreover, for \texttt{INTERP-down} and \texttt{INTERP-strange}, the probability distributions are transformed such that \emph{all} of the unlikely events (given the context) in the vocabulary are generally assigned more probability mass.
Thus, we expected \texttt{INTERP-down} and \texttt{INTERP-strange} models to yield high rates of infeasible events.
The sparse training set had a similar degenerative effect on the back-off models. With relatively few data points to inform high-order component models, the back-off model will successively probe low-order models until reaching unigrams---the base case.
Probing low-order models is not conducive to generating feasible events.
The purpose of smoothing techniques like interpolation and back-off is to use substantial contexts when possible before defaulting to lower-order models (e.g., unigrams and bigrams), but this is predicated on a training set that can support the estimation of higher-order models.

\section{Conclusion and Future Work}
\label{sec:concl}
We present a novel framework for automatic generation of actionable scenarios for Android apps. The framework uses two language models and three proposed flavors. For the evaluation we used several medium-large Android apps. On one hand, the results suggest that the {\tt BO} model is more suitable for generating unnatural scenarios with small and large corpora. On the other hand,  {\tt INTERP}  is not able to generate scenarios when the available corpora are small. If small corpora are available, {\tt INTERP} is more suitable for natural scenarios. Finally, the interactive mode in \monkeylab outperformed the serial mode in terms of coverage. 

We also provide a set of learned lessons for GUI-based testing of Android apps using actionable scenarios. The lessons recommend designing models for encapsulated components and increasing training corpus size of LMs. The lessons can be used by researchers to implement automated approaches for GUI-based testing that are more attractive and useful to mobile developers and testers. Actionable scenarios can also be combined with automatic generation of testing oracles~\cite{Lin:TSE14} for generating GUI-based test cases for mobile apps.

\balance
\bibliographystyle{abbrv}
\bibliography{ms}

\end{document}